# Toward a UTM-based Service Orchestration for UAVs in MEC-NFV Environment

Oussama Bekkouche[1], Miloud Bagaa[1], and Tarik Taleb[1,2]

[1]Communications and Networking Department, Aalto University, Finland. Email: firstname.lastname@aalto.fi
[2]Centre for Wireless Communications, University of Oulu, Finland. Email: tarik.taleb@oulu.fi

*Abstract*—The increased use of Unmanned Aerial Vehicles (UAVs) in numerous domains, will result in high traffic densities in the low-altitude airspace. Consequently, UAVs Traffic Management (UTM) systems that allow the integration of UAVs in the low-altitude airspace are gaining a lot of momentum. Furthermore, the $5^{th}$ generation of mobile networks (5G) will most likely provide the underlying support for UTM systems by providing connectivity to UAVs, enabling the control, tracking and communication with remote applications and services. However, UAVs may need to communicate with services with different communication Quality of Service (QoS) requirements, ranging form best-effort services to Ultra-Reliable Low-Latency Communications (URLLC) services. Indeed, 5G can ensure efficient Quality of Service (QoS) enhancements using new technologies, such as network slicing and Multi-access Edge Computing (MEC). In this context, Network Functions Virtualization (NFV) is considered as one of the pillars of 5G systems, by providing a QoS-aware Management and Orchestration (MANO) of softwarized services across cloud and MEC platforms. The MANO process of UAV's services can be enhanced further using the information provided by the UTM system, such as the UAVs' flight plans. In this paper, we propose an extended framework for the management and orchestration of UAVs' services in MECNFV environment by combining the functionalities provided by the MEC-NFV management and orchestration framework with the functionalities of a UTM system. Moreover, we propose an Integer Linear Programming (ILP) model of the placement scheme of our framework and we evaluate its performances. The obtained results demonstrate the effectiveness of the proposed solutions in achieving its design goals.

## I. INTRODUCTION

Motivated by the latest advances in robotics and communications technologies, the industry of Unmanned Aerial Vehicles (UAVs), commonly known as drones, knows a considerable booming. This is accompanied by intensive efforts from Aviation Safety Agencies (ASAs), such as the Federal Aviation Administration (FAA) and European Aviation Safety Agency (EASA) to address the variety of issues and novel challenges that face the integration of UAVs in the lowaltitude airspace. In this context, UAVs Traffic Management Systems (UTMs) are the mean by which ASAs will support the Urban Air Mobility (UAM) operations. Indeed, UTM systems provide a set of services that are vital for the safe and efficient operation of UAVs, such as flight planning, UAVs tracking, and intelligent flights deconfliction.

The $5^{th}$ generation of mobile communications (5G), will most likely provide the underlying support for UTM systems and UAVs' applications. Indeed, 5G is designed to accommodate a diversity of verticals and new use cases with different Quality of Service (QoS) requirements. This is mainly achieved by the introduction of the concept of Network Slicing (NS), that allows the creation and management of virtual network services with different communication requirements. Another important paradigm that was introduced in 5G, is Multi-Access Edge Computing (MEC), where computing and storage resources are pushed to the network edges (e.g., base stations, i.e., eNodeB or gNB) to host services with Ultra-Reliable and Low-Latency Communication (URLLC) requirements.

From UAVs perspective, NS can be used to allow optimized grouping of drones traffic in customized virtual network instances that ensure the QoS required by the UAVs' applications [1] (i.e., traffic control and use case related applications). Whereas, in addition to hosting traffic control services, MEC can be also used to offload intensive computation tasks from UAVs to the nearest edge cloud platform, harnessing the short response time guaranteed by MEC technology.

Network Function Virtualization (NFV), is considered as the main enabler of network slicing. Indeed, NFV allows the life cycle management and placement of Virtual Network Functions (VNFs) and the virtualized services over the NFV Infrastructure (NFVI), mainly hosted in the cloud. In a standard NFV implementation, the NFV Orchestrator (NFVO) is the component responsible for the placement and allocation of VNFs over the NFVI. The NFVO uses different algorithms and optimization models for selecting the best placement of VNFs, with different objectives, such as reducing the deployment cost or ensuring a specific QoS. However, the NFVO process, as it was originally defined by the European Telecommunication Standards Institute (ETSI), considers the NFVI as a cloud infrastructure hosted in remote operator's datacenters. Hence, most of the proposed solutions for VNFs placement doesn't take into account the placement of VNFs at the edge cloud infrastructure [2]. As a result, it would be impossible to find the adequate placements of VNFs that require low-latency and high reliability, which is the case of many UAVs applications and services.

To cope with this issue, ETSI has recently extended the functional architecture of MEC to enable the integration of MEC applications and services in the NFV environment [3]. Therefore, it would be possible to apply new placement schemes for different types of services with different QoS requirements, ranging from best-effort services to latency and reliability sensitive services, using the same service Management and Orchestration (MANO) framework. Another important aspect that must be taken into account during the placement of services, is the mobility of end users and the availability of services across the mobile network, this means that users should be able to access the virtualized services from any access point and regardless of their geographic locations.

In this work, we propose a new service-tailored placement scheme for UAVs' services (i.e., flights control and use case related services), harnessing the latest advances in UTM systems regulations and MEC-NFV standards. Indeed, the information that can be obtained from a UTM system such as the flight plan, the speed, and the service provided by the UAV can considerably enhance the MANO process to ensure an optimal latency and reliability aware service placement in MEC-NFV environment. Our paper has the following contributions. First, we propose an extended framework for the management and orchestration of UAVs' services in MECNFV environment by combining the functionalities provided by the MEC-NFV management and orchestration framework with the functionalities of a UTM system. Second, we propose an Integer Linear Programming (ILP) model for the NFVO process of the former framework, that aims at minimizing the deployment cost of services across the edge-cloud infrastructure while ensuring the required QoS in terms of latency, reliability, and bandwidth. The proposed model also provides the optimal routing information for the communication with UAVs during their flights.

The remainder of this paper is organized as follows. Section II presents the related works. Section III discusses our proposed framework. Our model for the NFVO process is detailed in Section IV. Section V presents the performances evaluation and results analysis of the proposed model. Finally, Section VI concludes the paper. It has to be noted that in the following sections, we use the terms VNF, service, and application interchangeably.

## II. RELATED WORK

### A. Integration of UAVs in 5G ecosystem

The work in [4] investigates the effect of communication latency and reliability on the Beyond Visual Line Of Sight (BVLOS) control of UAVs and proposes a new UTM architecture based on MEC, where the control services are hosted at the edge of the network in order to reduce the latency and the unreliability of communication with UAVs. A mathematical model and a heuristic for UAVs flights planning and MEC resources allocation are proposed in [5]. A 5G-based framework for preventive maintenance of critical infrastructure using UAVs is proposed in [6]. In [1], authors elaborate on how drones ecosystem can benefit from mobile technologies from LTE-Advanced to 5G, and summarize the key capabilities required by drones applications and the corresponding service requirements on mobile networks.

### B. Service orchestration in MEC-NFV environment

Doan-Van et al [7] have proposed a MEC framework, called APMEC, that interfaces with multiple NFV orchestrators to increase service availability. In [2], authors have proposed a novel service placement scheme tailored to URLLC in MECNFV environment. Work in [8] analyzed the compound effect of simultaneously considering virtual network functions and MEC applications deployed on the same network infrastructure and proposed a new architecture that aligns and integrates the MEC system with the MANO system. An NFV-based
MEC platform for efficient transmission of ultra-high quality multimedia is proposed in [9].

## III. PROPOSED FRAMEWORK

Fig. 1 depicts the interworking of the modules that compose our framework, ensuring an optimal placement of UAVs' services in terms of deployment cost, service availability and meeting QoS requirements. Each module will be detailed in the following subsections.

### A. The UTM module

As it was mentioned in section I, a UTM system is responsible for the safe operation of UAVs in the lowaltitude airspace using a set of federated services. The main services that can be identified in a typical UTM system are the registration and identification service, the flight planning service, the monitoring and tracking service, the restrictions management service, the command and control service, and the airspace authorization service. Hence, it has all the information regarding UAVs' flight plans (i.e., flights trajectories, flights speeds, starting and ending time of the flights), the targeted use case and the current state of each UAV that belongs to its management domain. These information, can be made available for an authorized third party such as a law enforcement agency. In our proposed framework, these information are shared with the network operator via the MCM module.

### B. Mobility and Communication Management (MCM)

The MCM module processes the information shared by the UTM system in order to extract the information related with the mobility of UAVs, the networking and computing resources consumed by the services required to successfully operate the UAVs, and the QoS required by these services. The data obtained from the UTM system can be processed by means of classical algorithmic, mathematical analysis or by applying Machine Learning (ML) techniques. The MCM module is composed of three sub-modules :

- *Resources Manager:* Extracts information about the amount of resources required by UAVs' services, such as the bandwidth, computing and storage resources. Such parameters can be deduced from the information related to the UAV's use case, for example, a UAV service that performs real-time video processing for object detection has a well known resources consumption. The resources information provided by this sub-module will be used by the NFVO during the resources allocation process.

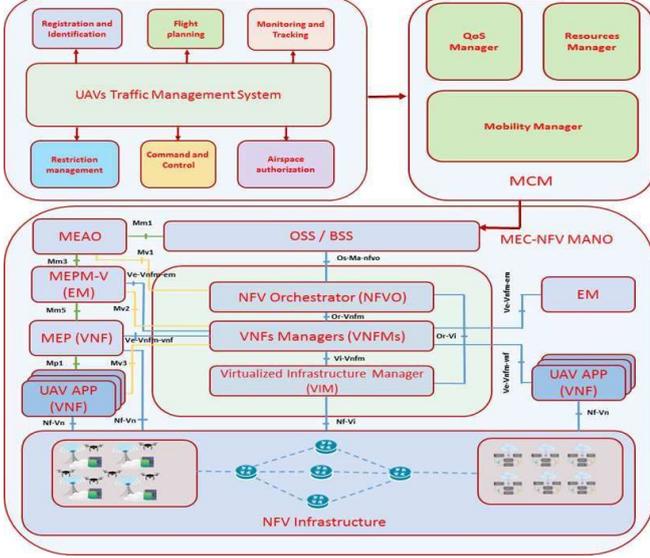

Fig. 1: UTM-based service orchestration for UAV in MECNFV environment.

- *QoS Manager:* Extracts information about the QoS required by UAVs' services, such as the latency and the reliability of the communication. Like the resources information, the QoS parameters can be deduced from the information related to the UAV's use case. The QoS information provided by this sub-module will be used by the NFVO during the placement process to select a host reachable via a communication path that ensures the specified QoS.
- *Mobility Manager:* Extracts information about the mobility of UAVs across the network. The mobility information is mainly represented by the list of base stations (i.e., eNodeB or gNB) that the UAVs will pass through during their flights, and the estimated arrival time to each base station. Such information can be deduced from the flight plans and the speed of UAVs, obtained from the UTM module. The hosting location of a given service should be accessible with respect to the QoS requirements, from any access point that belongs to the UAV trajectory.

## C. MEC-NFV Management and Orchestration (MEC-NFV MANO)

The MEC-NFV MANO module allows the orchestration of UAVs services across the NFVI composed of MEC infrastructures and remote cloud infrastructure. This is mainly achieved by deploying the main functional blocks of the MEC architecture defined in [10] as part of the NFV framework defined in [11]. A detailed description of MEC reference architecture in NFV environment can be found in [11].

## IV. NETWORK MODEL AND PROBLEM FORMULATION

### A. Network and System Model

As depicted in Fig. 2, we consider a mobile network that consists of three parts:
1) *Access Network*: A set of base stations that represent the access points, denoted by A.
2) *Transport and Core Network*: A network that connects the base stations to the operator's cloud datacenters. The set of nodes of this network is denoted by F.
3) *Cloud infrastructure*: A set of cloud platforms denoted by S.

Furthermore, as it is shown in Fig. 2, we consider the "Bump in the wire" [12] deployment of MEC, where MEC platforms are co-located at base stations, denoted by M, and at the aggregation points of the base stations, denoted by D. Let H denote the set of all possible hosts in the NFVI, that is, H = {S∪M∪D}. Each host $h \in H$ is associated with a resources capacity $R_h$ and a deployment cost per a time unit denoted by $C_h$.

We denote by $G(N,E,L,P,B)$ a weighted graph that represents all the NFVI, where the set of vertices $N$ = H∪A∪F, and the set of edges $E$ represents all the communication links in the NFVI. Each edge $(i,j) \in E$ is associated with a communication latency $l(i,j) \in L$, a failure probability $p(i,j) \in P$, and a bandwidth capacity $b(i,j) \in B$. Also, let $\eta(i)$ denote a function that return the list of neighbors of the node $i \in N$ in the graph $G$.

Let U denote the set of UAVs obtained from the UTM module, and $V_u$ the VNF associated with the UAV $u \in$ U. If a UAV is associated with more than one VNF, we simply replicate that UAV more than once in U. Each VNF $V_u$ is associated with a resources demand $D_u$, a bandwidth demand $B_u$, a reliability demand $P_u$, and a tolerated latency $L_u$. Moreover, each UAV $u \in$ U is associated with a flight trajectory $T_u$ that consists of the set of base stations that the UAV will fly across, that is, $T_u = \{a_1, a_2, ..., a_n\}$ with $a_i \in$ A. All these information are provided by the MCM module. Also, we define T, a set that symbolizes the discrete time periods.

We assume that the MCM module can predict the access point to which each UAV is connected at each time instant $t \in$ T using the flights plans information. Hence, we define the following constants:

$\forall u \in U, \forall a \in T_u, \forall t \in T:$

$$Z_{u a, t} = \begin{cases} 1 & \text{If the UAV } u \text{ is connected to the access point } a \text{ at time period } t. \\ 0 & \text{Otherwise.} \end{cases}$$

## B. Problem Formulation

In this subsection, we present our ILP model for the problem of placement of UAVs' VNFs in MEC-NFV environment. The proposed model aims at finding the optimal placement of a UAVs' VNF across the set of hosts H. In this context, an optimal placement scheme is the one that respects the communication requirements of the set of UAVs' VNFs, the capacities of NFVI hosts and the capacities of communication links, while minimizing the total deployment cost. We define the Boolean

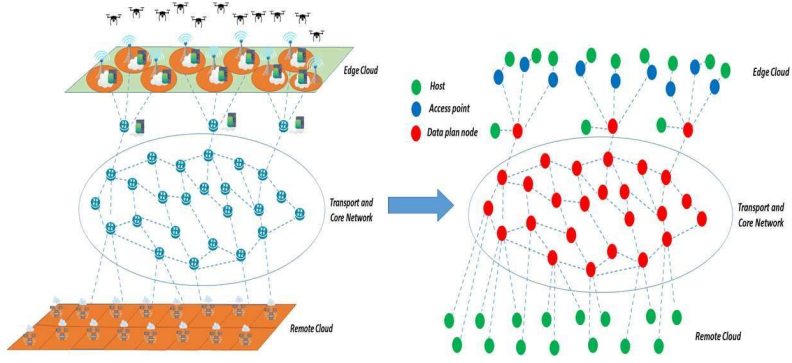

Fig. 2: Network Model.

variable $X_u^h$ that shows if an instance of the VNF $V_u$ is hosted in the host $h \in H$:

$\forall u \in U, \forall h \in H:$

$$\mathcal{X}_u^h = \begin{cases} 1 & \text{If the VNF } V_u \text{ is hosted in } h. \\ 0 & \text{Otherwise.} \end{cases}$$

In order to provide support for user mobility, in our model, we consider the possibility of reallocating the same UAV's VNF in multiple hosts, where each instance is responsible for serving the UAV when it is connected to different base stations during its flight. This will mainly enable URLLC between the UAVs and their VNFs, indeed, when a VNF requires a low-latency and high reliability communication with the UAV, it would be impossible to find a static location that satisfies this communication during the UAV flight, even at the edge of the network. This is why concepts like service reallocation, service replication, and service migration are vital for an optimal orchestration of services with URLLC requirements like UAVs' service. For this matter, we define the Boolean variable $K_{u,h,a}$ that shows if the instance of the VNF $V_u$ hosted in the host $h \in H$, serves when the UAV $u \in U$ is connected to the base station $a \in A$.

$\forall u \in U, \forall h \in H, \forall a \in T_u:$

$$K_{u,h,a} = \begin{cases} 1 & \text{If the VNF } V_u \text{ is hosted in } h \text{ and serves when} \\ & \text{the UAV } u \text{ is connected to the access} \\ & \text{point } a. \\ 0 & \text{Otherwise.} \end{cases}$$

Also, we define the variable $\mathcal{Y}_{u,h,a}^{i,j}$ that shows if the instance of the VNF $V_u$ hosted in the host $h \in H$, use the link $(i,j) \in E$ in the communication path with the UAV $u \in U$ when it is connected to access point $a \in A$.

$\forall u \in U, \forall h \in H, \forall a \in T_u, \forall i \in N, \forall j \in \eta(i)$

$$Y_{u,h,a,i,j} = \begin{cases} 1 & \text{If the VNF } V_u \text{ is hosted in } h \text{ and use} \\ & \text{the link } (i,j) \text{ in the communication} \\ & \text{path with} \\ & \text{the UAV } u \text{ when it is connected access} \\ & \text{point } a. \\ 0 & \text{Otherwise.} \end{cases}$$

The objective function of our ILP, that aims at minimizing the total deployment cost can be expressed as follows:

$$\min \sum_{h \in H} \sum_{t \in T} \sum_{u \in U} \sum_{a \in T_u} K_{u,h,a} \times Z_{ua,t} \times D_u \times C_h$$

The former objective function is subject to the following sets of constraints:

*1) Placement constraints:* The constraints defined in 1, ensure that all UAVs' VNFs will we placed in at least one host (i.e., one instance of the VNFs will be placed in a host).

$$\forall u \in U : \sum_{h \in H} X_u^h \geq 1 \quad (1)$$

Whereas, constraints defined in 2, ensure that an instance of the VNF $V_u$ that is not hosted in a host $h$ will not serve the UAV whatever the connecting base station is.

$$\forall u \in U, \forall h \in H, \forall a \in T_u : K_{u,h,a} \leq X_u^h \quad (2)$$

Constraints defined in 3 ensure that no instance of the VNF $V_u$ will be placed in the host $h \in H$ if this instance will not serve the UAV during its flight.

$$\forall u \in U, \forall h \in H : X_u^h \leq \sum K_{u,h,a} \quad (3)$$

Constraints defined in 4 ensure that each UAV $u \in U$ will be served by exactly one VNF instance during its flight.

$$\forall u \in U, \forall a \in T_u : \sum_{h \in H} K_{u,h,a} = 1 \quad (4)$$

*2) Hosts capacities constraints:* The constraints, defined in 5, ensure that the amount of resources consumed by the set of VNFs hosted in a host $h \in H$, at any time period $t \in T$ doesn't exceed the resources capacity of that host.

$$\forall h \in H, \forall t \in T : \sum_{u \in U} \sum_{a \in T_u} K_{u,h,a} \times Z_u^{a,t} \times D_u \leq R_h \quad (5)$$

*3) Links capacities constraints:* The constraints, defined in 6, ensure that the amount of traffic that passes thought any communication link $(i,j) \in E$ at any time period $t \in T$ doesn't exceed the bandwidth of that link. $\forall i \in N, \forall j \in \eta(i), \forall t \in T :$

$$\sum_{u \in U} \sum_{h \in H} \sum_{a \in T_u} Y_{u,h,a}^{i,j} \times Z_u^{a,t} \times B_u \leq b(i,j) \quad (6)$$

*4) Routing constraints:* Constraints, defined in 7, 8, 9, 10 and 11, ensure the establishment of a communication path without loop and ramifications, between the instance of the VNF $V_u$ hosted in $h \in H$ and the UAV $u$ when it is connected to the base station $a$

$$\forall u \in U, \forall h \in H, \forall a \in T_u, \forall i \in N, \forall j \in \eta(i) : Y_{u,h,a}^{i,j} \leq K_{u,h,a} \quad (7)$$

$$\forall u \in \mathcal{U}, \forall h \in \mathcal{H}, \forall a \in T_u : \sum_{j \in \eta(h)} \mathcal{Y}_{u,h,a}^{h,j} = \mathcal{K}_{u,h,a} \quad (8)$$

$$\forall u \in \mathcal{U}, \forall h \in \mathcal{H}, \forall a \in T_u : \sum_{i \in \eta(a)} \mathcal{Y}_{u,h,a}^{i,a} = \mathcal{K}_{u,h,a} \quad (9)$$

$$\forall u \in \mathcal{U}, \forall h \in \mathcal{H}, \forall a \in T_u : \sum_{j \in \eta(a)} \mathcal{Y}_{u,h,a}^{a,j} = 0 \quad (10)$$

$\forall u \in U, \forall h \in H, \forall a \in T_u, \forall i \in N \setminus \{a,h\} :$

$$\sum_{j \in \eta(i)} Y_{u,h,a}^{i,j} = \sum_{j \in \eta(i)} Y_{u,h,a}^{j,i} \quad (11)$$

*5) Latency constraints:* Constraints, defined in 12, ensure that the established path between the instance of the VNF $V_u$ hosted in the host $h$ and the UAV $u$ when it is connected to the base station $a$, has a latency less or equal to the latency tolerated by $V_u$.

$$\forall u \in \mathcal{U}, \forall h \in \mathcal{H}, \forall a \in T_u : \sum_{i \in N} \sum_{j \in \eta(i)} \mathcal{Y}_{u,h,a}^{i,j} \times l(i,j) \leq \mathcal{L}_u \quad (12)$$

*6) Reliability constraints:* Given a communication path composed of $n$ links, with $\{p_1, p_2, \dots, p_n\}$ representing the failure probability of these links, the survivability (i.e., success) probability of the full path P, is given as follows:

$$P = \prod_{i=1}^{n}(1 - p_i) \quad (13)$$

As a result, the reliability constraints of the established path between the instance of the VNF $V_u$ hosted in the host $h$ and the UAV $u$ when it is connected to the base station $a$ can be expressed as follows:

$$\forall u \in U, \forall h \in H, \forall a \in T_u : \prod_{(i,j) \in E}\left(1 - p(i,j) \times Y_{u,h,a}^{i,j}\right) \geq P_u \quad (14)$$

However, constraint 14 is not linear, so, we perform the following transformations in order to linearize it.

$$\log\left[\prod_{(i,j) \in E}(1 - p(i,j) \times \mathcal{Y}_{u,h,a}^{i,j})\right] \geq \log(\mathcal{P}_u) \quad (15)$$

$$\sum_{(i,j) \in E} \log(1 - p(i,j) \times \mathcal{Y}_{u,h,a}^{i,j}) \geq \log(\mathcal{P}_u) \quad (16)$$

$$\sum_{(i,j) \in E} Y_{u,h,a}^{i,j} \times \log(1 - p(i,j)) \geq \log(P_u) \quad (17)$$

$$\sum_{i \in N} \sum_{j \in \eta(i)} Y_{u,h,a}^{i,j} \times \log(1 - p(i,j)) \geq \log(P_u) \quad (18)$$

After introducing all constraints and their respective transformations, the final ILP model to optimize is:

$$\min \sum_{h \in \mathcal{H}} \sum_{t \in \mathcal{T}} \sum_{u \in \mathcal{U}} \sum_{a \in T_u} \mathcal{K}_{u,h,a} \times \mathcal{Z}_u^{a,t} \times D_u \times \mathcal{C}_h$$

s.t. $\begin{cases} Constraints & 1-12 \\ Constraints & 18 \end{cases}$

## V. EVALUATION

In this section, we evaluate our solution in terms of: *i)* the run time complexity; *ii)* the deployment cost; *iii)* the number of replicates of a VNF; *iv)* the ability to provide a latency and reliability aware service placement in MEC-NFV environment.

We evaluated the former metrics by varying the number of UAVs in the networks (i.e., the number of VNFs). For each scenario, we run 30 repetitions, altering the UAVs' VNFs resources demand, bandwidth demand, reliability demand, and the tolerated latency. Moreover, in each iteration, we consider a new network topology where we vary the parameters according to the values shown in Table I. It has to be noted that we have considered that the cost of the deployment of services in edge hosts is higher than the cost of the deployment in cloud hosts, as the amount of resources available at edge hosts is much less than cloud hosts.

TABLE I: Simulation parameters.

| UAVs' parameters | Values |
| --- | --- |
| UAVs Number | 5, 10, 15, 20, 25 |
| Resources demand | 10 - 20 |
| Bandwidth demand | 50 - 100 Mbps |
| Reliability demand | 0.95 - 0.99 |
| Tolerated latency | 1 - 50 ms |

| Network parameters | Values |
|---|---|
| Base stations number | 20 |
| Number of edge hosts (i.e., |M|) | 20 |
| Edge hosts resources capacities | 200 - 400 |
| Edge hosts cost | 500 - 1000 |
| Number of aggregation hosts | 3 |
| Aggregation hosts resources capacities | 400 - 800 |
| Aggregation hosts cost | 250 - 500 |
| Number of cloud hosts | 15 |
| Cloud hosts resources capacities | 800 - 1600 |
| Cloud hosts cost | 100 - 300 |
| Core network nodes number | 37 |
| Links bandwidth | 1 - 10 Gbps |
| Links failure probability | 0 - 0.01 |
| Links latency | 1 - 3 ms |
| Time periods | Value |
| T | {1,2, ... 30 } |

Fig. 3(a) shows the impact of the number of UAVs on the cost of proposed solution. From this figure, we observe that the cost increases proportionally with the number of UAVs in the network. The more UAVs in the network are, the more the number of services to be deployed is. Meanwhile, Fig. 3(b) shows the impact of the number of UAVs on the execution time. This figure shows that the number of UAVs has a negative impact on the execution time. Formally, increasing the number of UAVs would lead to increase the number of variables and constraints in our ILP, and hence, would have a negative impact on the complexity of proposed solution. From this figure, we observe that our solution could take up to 2500 seconds to provide the optimal placement of UAVs' VNFs, which implies that the solution must be executed offline before starting UAVs' missions.

Fig. 3(c) shows the impact of the mission length represented by the number of base stations in the UAVs' trajectories, on the average number of VNFs replicates that must be deployed in the network in order to ensure the desired QoS during the flight of the UAV. Indeed, increasing the mission length will increase the possibility that the UAV moves away from its original serving VNF, and hence the probability that the end-to-end latency to be unsatisfied, which requires the reallocation of the VNF in another host. For this reason, as depicted in Fig. 3(c), increasing the mission length will increase the number of replicates that should be deployed in the network to satisfy the end-to-end latency. We observe that whatever the length of the mission, our solution succeeds to ensure the tolerated end-to-end latency while deploying in the worst case scenarios 2.3 replicates for each VNF in average.

Finally, the scatter graph depicted in Fig. 3(d), shows the type of hosts selected to host the UAVs' VNFs according to the latency and reliability demand of this latter. We notice that when the tolerated latency varies between 1ms and 3ms, the services are placed at the edge hosts co-located at base stations, regardless of the value of the required reliability. Nevertheless, when the tolerated latency varies between 4ms and 14ms, the services are placed at the edge hosts co-located at the aggregation nodes, also, regardless of the value of the required reliability. On the other hand, when he tolerated latency varies between 15ms and 50ms, we can distinct two cases. The first case is when the required reliability varies between 0.95 and 0.972, we notice that the UAVs' VNFs are hosted in the remote cloud infrastructure. The second case, is when the required reliability varies between 0.973 and 0.99, we notice that the UAVs' VNFs are hosted in the edge hosts co-located at the aggregation nodes.

## VI. CONCLUSION

In this paper, we introduced a new aligned process for the orchestration of UAVs services in MEC-NFV environment, proposing a framework that combines the functionalities of UTM and MANO systems for optimal placement and provisioning of UAVs' services. The proposed framework can ensure the deployment of UAVs' service across the infrastructure composed of MEC platforms co-located at base stations and cloud platforms hosted in the operator's remote datacenters, while ensuring that the services will be available and accessible with respect to QoS requirements, along the flight trajectory of UAVs. Moreover, we proposed an optimization model for the aligned orchestration process that aims at minimizing the total deployment cost. The simulation results demonstrated the effectiveness of the proposed solution in achieving its design goals.

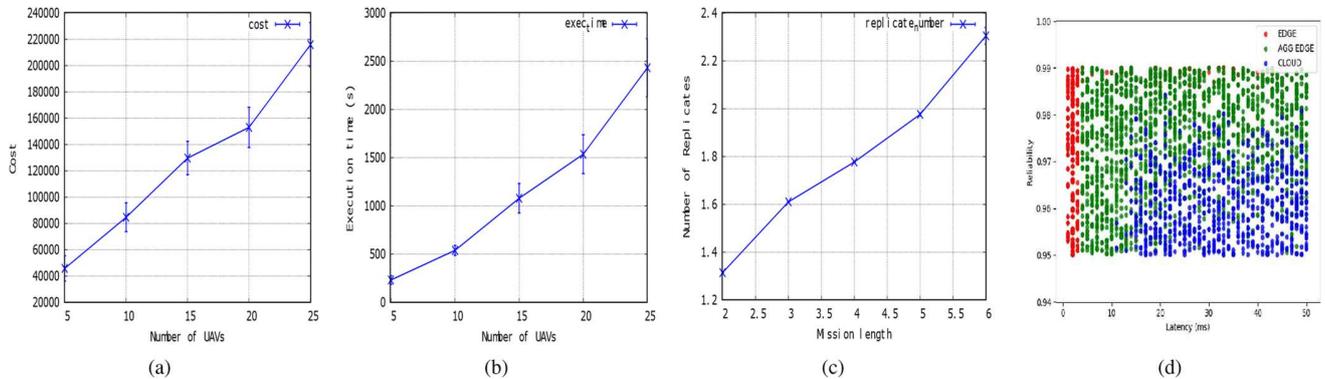

Fig. 3: Performances evaluation.

ACKNOWLEDGMENT

This work is partially supported by the European Union's Horizon 2020 research and innovation program, under the 5G!Drones project with grant agreement No. 857031, and Primo-5G project with grant agreement No. 815191. It is also supported in part by the 6G Flagship project with grant agreement No. 318927, and by the Academy of Finland under CSN project with grant No. 311654.